\documentclass[reprint,superscriptaddress,amsmath,amssymb,aps,prl]{revtex4-2}
\usepackage{graphicx}
\usepackage{dcolumn}
\usepackage{bm}
\begin{document}
\title{Pomeranchuk Effect and Tunable Quantum Phase Transitions in 3L-MoTe$_2$/WSe$_2$}

\author{Mingjie Zhang}
\affiliation{Beijing National Laboratory for Condensed Matter Physics, Institute of Physics, Chinese Academy of Sciences, Beijing, China}
\affiliation{School of Physical Sciences, University of Chinese Academy of Sciences, Beijing, China}

\author{Xuan Zhao}
\affiliation{Beijing National Laboratory for Condensed Matter Physics, Institute of Physics, Chinese Academy of Sciences, Beijing, China}
\affiliation{School of Physical Sciences, University of Chinese Academy of Sciences, Beijing, China}

\author{Kenji Watanabe}
\affiliation{Research Center for Functional Materials, National Institute for Materials Science, Tsukuba, Japan}

\author{Takashi Taniguchi}
\affiliation{International Center for Materials Nanoarchitectonics, National Institute for Materials Science, Tsukuba, Japan}

\author{Zheng Zhu}
\affiliation{Kavli Institute for Theoretical Sciences, University of Chinese Academy of Sciences, Beijing, China}

\author{Fengcheng Wu}
\affiliation{School of Physics and Technology, Wuhan University, Wuhan, China}

\author{Yongqing Li}
\affiliation{Beijing National Laboratory for Condensed Matter Physics, Institute of Physics, Chinese Academy of Sciences, Beijing, China}
\affiliation{School of Physical Sciences, University of Chinese Academy of Sciences, Beijing, China}

\author{Yang Xu}
\email[]{yang.xu@iphy.ac.cn}
\affiliation{Beijing National Laboratory for Condensed Matter Physics, Institute of Physics, Chinese Academy of Sciences, Beijing, China}



\begin{abstract}
Many sought-after exotic states of matter are known to emerge close to quantum phase transitions, such as quantum spin liquids (QSL) and unconventional superconductivity. It is thus desirable to experimentally explore systems that can be continuously tuned across these transitions. Here, we demonstrate such tunability and the electronic correlation effects in triangular moiré superlattices formed between trilayer MoTe$_2$ and monolayer WSe$_2$ (3L-MoTe$_2$/WSe$_2$). Through electric transport measurements, we firmly establish the Pomeranchuk effect observed at half filling of the first moiré subband, where increasing temperature paradoxically enhances charge localization. The system simultaneously exhibits the characteristic of a Fermi liquid with strongly renormalized effective mass, suggesting a correlated metal state. The state is highly susceptible to out-of-plane electric and magnetic fields, which induce a Lifshitz transition and a metal-insulator transition (MIT), respectively. It enables identification of a tricritical point in the quantum phase diagram at the base temperature. We explain the Lifshitz transition in terms of interlayer charge transfer by applying the vertical electric field, which leads to the emergence of a new Fermi surface and immediate suppression of the Pomeranchuk effect. The existence of quantum criticality in the magnetic filed induced MIT is supported by scaling behaviors of the resistance. Our work shows the 3L-MoTe$_2$/WSe$_2$ lies in the vicinity to the MIT point of the triangular lattice Hubbard model, rendering it a unique system to manifest the rich correlation effects at an intermediate interaction strength.

\end{abstract}


\maketitle
The discovery of correlated phenomena in twisted bilayer graphene opens a new avenue for studying electron correlations in two-dimensional (2D) moiré superlattices, which host novel band structure reconstruction and exotic quantum phases \cite{cao2018correlated, cao2018unconventional, balents2020superconductivity,andrei2021marvels}. Lately, semiconductor moiré systems based on transition-metal dichalcogenide (TMDC) heterostructures have been shown to offer alternative platforms with appealing opportunities \cite{wu2018hubbard, wu2019topological}. Moiré excitons/trions/polaritons \cite{alexeev2019resonantly, jin2019observation, seyler2019signatures, tran2019evidence, liu2021signatures, wang2021moire, zhang2021van}, Mott insulators \cite{tang2020simulation, wang2020correlated, shimazaki2020strongly}, generalized Wigner crystals at fractional fillings \cite{regan2020mott, xu2020correlated, huang2021correlated}(including those spontaneously break rotational symmetry \cite{jin2021stripe, li2021imaging}), and topological phases \cite{Li2021} have been experimentally observed. It’s generally accepted that the low energy Hamiltonian of the TMDC moiré system can be effectively described by the triangular lattice Hubbard model \cite{wu2018hubbard}, which has been one of the central paradigms in describing correlated systems with geometrical frustration \cite{yoshioka2009quantum, yang2010effective, shirakawa2017ground, szasz2020chiral, wietek2021mott}. When doped with one charge per site (half-filled band), the large spin entropy owing to the frustration and competing orders can give rise to the Pomeranchuk effect or an exotic nonmagnetic insulating state (e.g. QSL), depending on the interaction strength $U/t$, where $U$ is the on-site Coulomb repulsion energy and $t$ is the hopping integral \cite{wietek2021mott}. Despite extensive research efforts on the triangular lattice Hubbard model, many aspects of its rich phase diagram remain unexplored due to high complexity of the problem and the lack of suitable experimental platforms in different coupling regimes.

In TMDC moiré homo- or hetero- bilayers (such as twisted WSe$_2$ or MoTe$_2$/WSe$_2$) with small band offsets between neighboring layers, the charge density and vertical electric fields can be independently controlled by electrostatic gates, enabling an in-situ access to a filling or a bandwidth tuned phase diagram \cite{li2021continuous, ghiotto2021quantum}. Utilizing the sensitivity of the interaction strength to the dielectric environment and interlayer coupling, we fabricate a new type of moiré heterostructure consisting of trilayer MoTe$_2$ and monolayer WSe$_2$. The multilayer TMDC (3L-MoTe$_2$ here) based superlattices can also potentially offer new opportunities to explore moiré physics (e.g. that arises from the $\Gamma$ valley instead of the $K$ valley) exhibiting very different symmetries and strengths of spin-orbital coupling \cite{Angeli2021}. 

\begin{figure*}[t]
	\centering
	\includegraphics[scale=1]{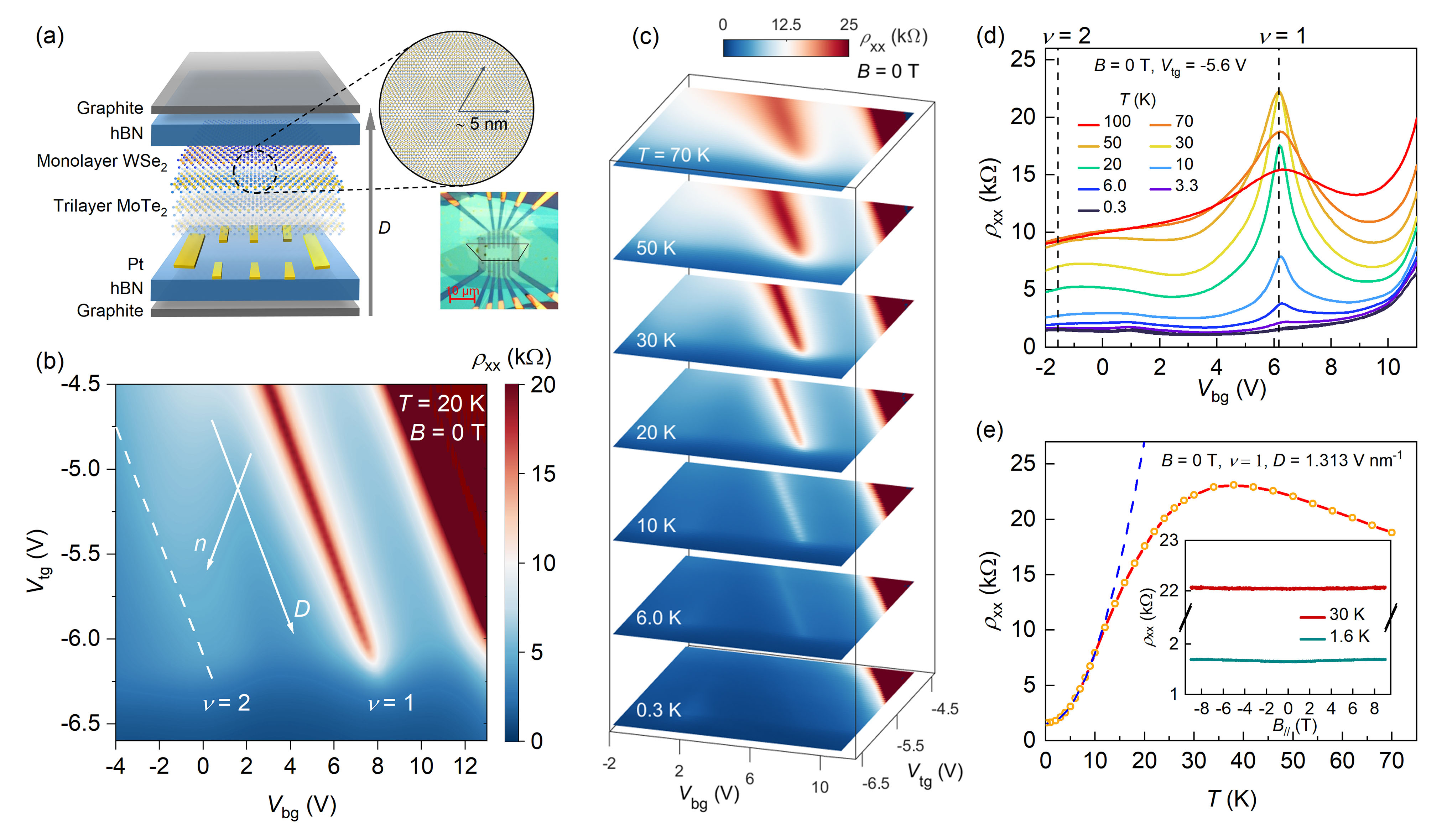}
	\caption{Device structure and Pomeranchuk effect in 3L-MoTe$_2$/WSe$_2$ moiré superlattices. (a) Schematic (left) and optical micrograph (lower right) of the device with double graphite gates. The conduction channel of the device is outlined by the black curves in the micrograph. The scale bar is 10 $\mu$m. (b) 2D longitudinal resistance $\rho_{xx}$ as a function of top and bottom gate voltages ($V_{tg}$ and $V_{bg}$, respectively) at $T$ = 20 K and zero magnetic field. An obvious resistivity peak is observed at half filling $\nu=1$. The dashed line marks the expected position of full filling $\nu=2$. The two arrows denote the directions of increasing $D$ and $n$, respectively. (c) Evolution of $\rho_{xx}$ as a function of $T$. Each color map plane shows the $\rho_{xx}$ versus $V_{tg}$ and $V_{bg}$ at a fixed $T$. (d) The $\rho_{xx}$ as a function of bottom gate voltage at a fixed top gate voltage $V_{tg}=-5.6$ V at different temperatures. Two dashed lines indicate the position of $\nu=1$ and $\nu=2$, respectively. (e) Temperature dependence and in-plane magnetic field dependence (inset) of $\rho_{xx}$ at $\nu=1$ and $D=1.313$ V nm$^{-1}$. At low temperatures ($T < 10$ K), $\rho_{xx}(T)$ follows a Fermi liquid behavior $AT^2+\rho_{0}$ (blue dashed curve).}
\end{figure*}

In this letter, we discover a rich and exotic quantum phase diagram tuned by both out-of-plane electrical and magnetic fields in the 3L-MoTe$_2$/WSe$_2$. At half filling of the first moiré subband, we find a resistance peak that counterintuitively develops with increasing temperature. This behavior resembles the Pomeranchuk effect observed in He-3 and is attributed to the close vicinity of our system to the MIT point in the triangular lattice Hubbard model. Due to small valence band offset of the two materials, the hole-type charge carriers are first injected into the MoTe$_2$ and the subband population from WSe$_2$ can be continuously tuned by the vertical electric field, which drives a Lifshitz transition and alters the charge transport greatly. The process is accompanied by breakdown of the single-band Hubbard model description of the system. The magnetic field, which suppresses spin fluctuations, can induce a MIT following the quantum critical scaling. We observe continuous closure of the charge gap when approaching the quantum critical point from the insulating side and divergence of effective quasiparticle mass from the metallic side.

Natural trilayer MoTe$_2$ (2H phase) and monolayer WSe$_2$ flakes are mechanically exfoliated and angle-aligned to form 3L-MoTe$_2$/WSe$_2$ moiré superlattices (see supplementary materials for more details). The two materials have a lattice mismatch about 7\%, generating a maximum moiré wavelength ~5 nm and a superlattice density $\sim$5$\times$10$^{12}$ cm$^{-2}$. The 3L-MoTe$_2$/WSe$_2$ heterostructure is encapsulated by hexagonal boron nitride sheets (about 5 to 10 nm thick) as gate dielectrics and patterned into a quasi-Hall bar structure (see the device schematic and an optical image in Fig. 1a) for electronic transport measurements. Top and bottom gate electrodes are made by few-layer graphite. Charge carriers with density $n$ (hole type) and a vertical electric displacement field $D$ can be introduced by the combination of top and back gates ($V_{tg}$ and $V_{bg}$, respectively).

\begin{figure*}[t]
	\centering
	\includegraphics[scale=1]{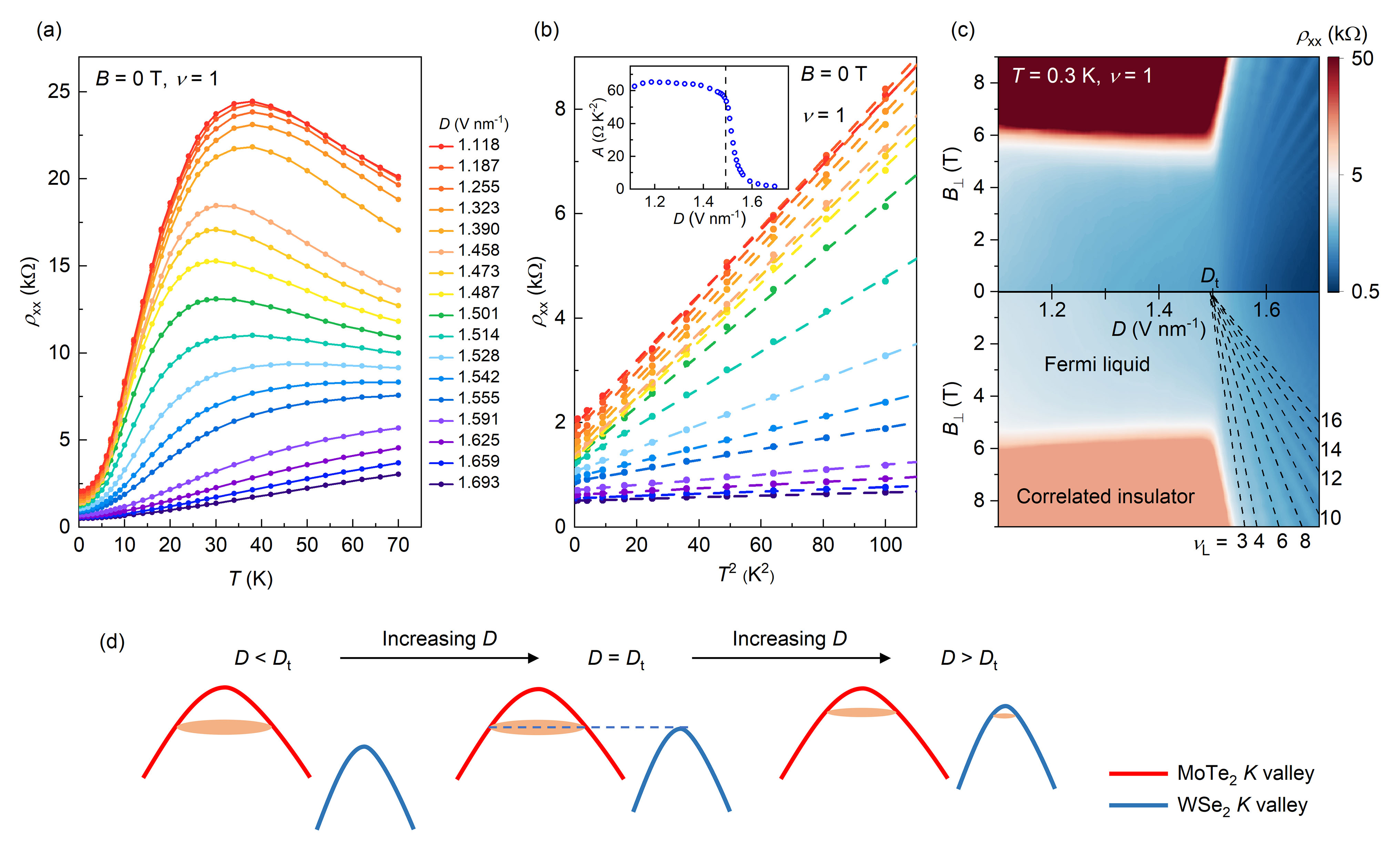}
	\caption{Electric displacement field driven Lifshitz transition. (a) Temperature dependences of $\rho_{xx}$ with varying displacement field $D$ at $\nu=1$ and $B=0$ T. (b) Same data as (a) (sharing the same color coding) with x-axis scaled in $T^2$ below $\sim$10 K. The dashed lines are the corresponding linear fits. Inset: Fitting parameter $A$ as a function of $D$. The vertical dashed line indicates the inflection point at $D=D_t$. (c) Upper diagram shows $\rho_{xx}$ as a function of out-of-plane magnetic field $B_{\perp}$ and $D$ at $\nu$ = 1 and $T$ = 0.3 K. Lower diagram shows color-coded regions representing different states. The Landau fan (linear dashed lines) converges to $D_t$, with filling factors $\nu_L$ labelled accordingly. (d) Schematics showing the Lifshitz transition and evolution of band alignments (\emph{K} valleys of MoTe$_2$ and WSe$_2$) with increasing electric displacement field $D$. Various effects (such as moiré band formation and interaction-induced mass renormalization) are neglected in this simple schematic. Above $D_t$, charges are transferred between the MoTe$_2$ and the WSe$_2$.}
\end{figure*}

Fig. 1b shows the longitudinal 2D sheet resistance $\rho_{xx}$ as a function of $V_{tg}$ and $V_{bg}$ measured at an intermediate temperature $T$ = 20 K. The two arrows denote the directions for increasing $D$ or $n$, respectively. Notably, the sample exhibits a resistance peak along a constant density line (with $n$ $\approx5\times10^{12}$ cm$^{-2}$, see Supplementary Figure 1d), corresponding to one hole per moiré cell or half filling of the first moiré subband, namely moiré filling factor $\nu=1$. Upon increasing the $D$ field, the peak disappears abruptly (near $D$ $\sim$ 1.5 V nm$^{-1}$). Except for the absence of a $\nu=2$ band insulating state, the behavior is reminiscent of a recently reported bandwidth-tuned Mott transition in MoTe$_2$/WSe$_2$ heterobilayers \cite{li2021continuous}. However, we find that the resistance peak here is gradually suppressed with decreasing temperature and completely vanishes at the lowest $T = 0.3$ K, in sharp contrast to the Mott insulating state \cite{tang2020simulation, wang2020correlated, regan2020mott}(see the resistance color mappings and constant $V_{tg}$ linecuts at selective temperatures in Fig. 1c and 1d, respectively). In other words, the resistance peak at half filling only gets enhanced at elevated temperatures, indicating a heat induced charge localization at the specific commensurate doping level $\nu=1$. Similar phenomenon in moiré superlattices was first reported in magic-angle twisted bilayer graphene (TBG) \cite{rozen2021entropic, saito2021isospin} and referred as the Pomeranchuk effect, which is an electronic analogue of the higher-temperature solidification of He-3 \cite{pomeranchuk1950theory}. This unusual behavior is due to the larger spin entropy in the localized state (solid phase in He-3 or singly occupied state in 3L-MoTe$_2$/WSe$_2$). The system is then favorable to increase the degree of localization upon heating. We also note that the neighboring local moments in 3L-MoTe$_2$/WSe$_2$ have antiferromagnetic superexchange interactions, whereas in TBG isospin polarization are favored for the $|\nu|=1$ states \cite{rozen2021entropic, saito2021isospin}.

A representative $\rho_{xx}$ versus $T$ curve for the $\nu=1$ state (at $D$ = 1.313 V nm$^{-1}$) is shown in Fig. 1e. The sample exhibits an insulating behavior (d$\rho_{xx}$/d$T<0$) at temperatures higher than $\sim$40 K, below which a crossover to a metallic behavior (d$\rho_{xx}$/d$T>0$) is observed with $\rho_{xx}$ decreasing by more than one order of magnitude (from $\sim$40 K to 0.3 K). The maximum of $\rho_{xx}$ is nearly twice the value of the Ioffe-Regel criteria $R_{c}=h/2e^2\approx$ 12.9 k$\Omega$ for the MIT of a 2D electron gas with double degeneracy. At $T<\sim$10 K (where $\rho_{xx}$ reaches $\sim$$R_{c}$), the $\rho_{xx}$($T$) can be well described by a Fermi liquid behavior $\rho_{xx}(T)=AT^2+\rho_{0}$ (highlighted by the dashed blue curve), where $A^{1/2}$ is proportional to the quasiparticle effective mass $m^*$ (Kadowaki-Woods scaling) \cite{rice1968electron, kadowaki1986universal} and $\rho$$_{0}$ is the residual resistance. The value of $A$ is extracted to be $\sim$64 $\Omega$/K$^2$ or $\sim$$4.5$ $\mu\Omega$ cm/K$^2$, considering a thickness of $\sim$0.7 nm for a single layer MoTe$_2$. This value is comparable to those of many heavy fermion systems, indicating substantial quasiparticle-quasiparticle scatterings and a strongly renormalized effective mass at low temperatures \cite{jacko2009unified}. In the inset of Fig. 1e, we plot the in-plane magnetic field ($B_{||}$) dependences of $\rho_{xx}$ at two different temperatures ($T$ = 1.6 K and 30 K). Unlike the TBG at $|\nu|=1$ \cite{rozen2021entropic, saito2021isospin}, 3L-MoTe$_2$/WSe$_2$ is insensitive to the in-plane magnetic field.

Now we discuss the effect of the vertical displacement field ($D$) at $\nu$ = 1. At different temperatures, a similar value of $D$ field is observed where the resistance maxima disappear (Fig. 1c). In Fig. 2a, we plot the temperature dependence (0.3 K to 70 K) of $\rho_{xx}$ at different $D$ fields. The $\rho_{xx}$ gradually changes from a non-monotonic behavior to a monotonic and metallic behavior with a weaker $T$-dependence at larger $D$ fields. At low temperatures ($T<\sim$$10$ K), all resistance curves follow $T^2$ dependences with $\rho_{xx}(T)=AT^2+\rho_{0}$ (Fig. 2b). The extracted $A$ coefficient is plotted in the inset of Fig. 2b as a function of the $D$ field. The $A$ value stays nearly a constant at small $D$ fields while decreases drastically by more than 30 folds starting from an inflection point at $\sim$1.49 V nm$^{-1}$, which is denoted as the transition field $D_t$ (dashed vertical line in the inset of Fig. 2b). It indicates strong suppression of electronic interactions in the system and an $\sim$80\% reduction of effective quasiparticle mass. The high-temperature value of $\rho_{xx}$ near $D_t$ is close to $R_{c}\approx$ 12.9 k$\Omega$. On both sides of the transition, the sample exhibits metallic ground states with Fermi liquid behaviors. This “metal-metal” transition observed here contrasts the electric field induced MIT found in MoTe$_2$/WSe$_2$ \cite{li2021continuous}. The $D_t$ value (corresponding to a $E_t=D_t/\varepsilon_r=0.50$ V nm$^{-1}$, where $\varepsilon$$_r$ is the out-of-plane dielectric constant of hBN) is also smaller than the critical electric field (with $E_c=0.65$ V nm$^{-1}$) for MIT in the heterobilayer \cite{li2021continuous}.

The magnetotransport studies (under out-of-plane magnetic fields $B_\perp$) suggest that the $D$ field in our sample introduces a Lifshitz transition at $D_t$, beyond which charges are redistributed between different subbands/layers. As shown in Fig. 2c, the $\rho_{xx}$ ($\nu$ = 1) is plotted as a function of both $D$ and $B_\perp$ at the base temperature $T$ = 0.3 K. At small values of $D$, the sample encounters a transition to a correlated insulating state (with $\rho_{xx}>10^7$ $\Omega$, see more data in Fig. 3) above $\sim$6 T. Combined with the barely changed in-plane magnetoresistance (inset of Fig. 1e), the strong magnetic anisotropy suggests an Ising type of spin-orbit coupling of the charge carriers that likely arise from \emph{K} valleys of the MoTe$_2$. It is also consistent with the fact that the finite out-of-plane displacement field lifts the layer degeneracy and leads to a strong localization of \emph{K} valley wavefunctions on the topmost layer of the trilayer MoTe$_2$ \cite{movva2018tunable}. 

At larger values of $D$, clear Shubnikov–de Hass (SdH) oscillations due to the formation of Landau levels are observed at $B_{\perp}>\sim$3 T. The generated Landau fan (also indicated by the dashed lines in the lower diagram) converges to $D$ = 1.492 V nm$^{-1}$ in the zero $B_{\perp}$ field limit. The value coincides (while being more accurate) with the transition field $D_t$ discussed above and indicates the emergence of a well-defined Fermi surface at $D>D_t$. Hence it corresponds to a Lifshitz transition. The linear dependences of the Landau levels on the $D$ field indicates the size of the new Fermi surface is directly proportional to ($D-D_t$). A cyclotron effective mass $\sim$0.5$m_e$ (with $m_e$ being the free electron mass) can be determined from analyzing the temperature dependence of the SdH oscillations (see more details in Supplementary Figure 3). It is close to the effective mass found in monolayer WSe$_2$ \emph{K} valleys \cite{movva2017density, gustafsson2018ambipolar}. We thus conclude that the charge transfer induced by the vertical electric field happens between the \emph{K} valleys of the MoTe$_2$ top layer and the adjacent WSe$_2$ monolayer (schematics shown in Fig. 2d), giving rise to two sets of quantum oscillations observed at $D>D_t$ (see Supplementary Figure 4).

\begin{figure*}[t]
	\centering
	\includegraphics[scale=1]{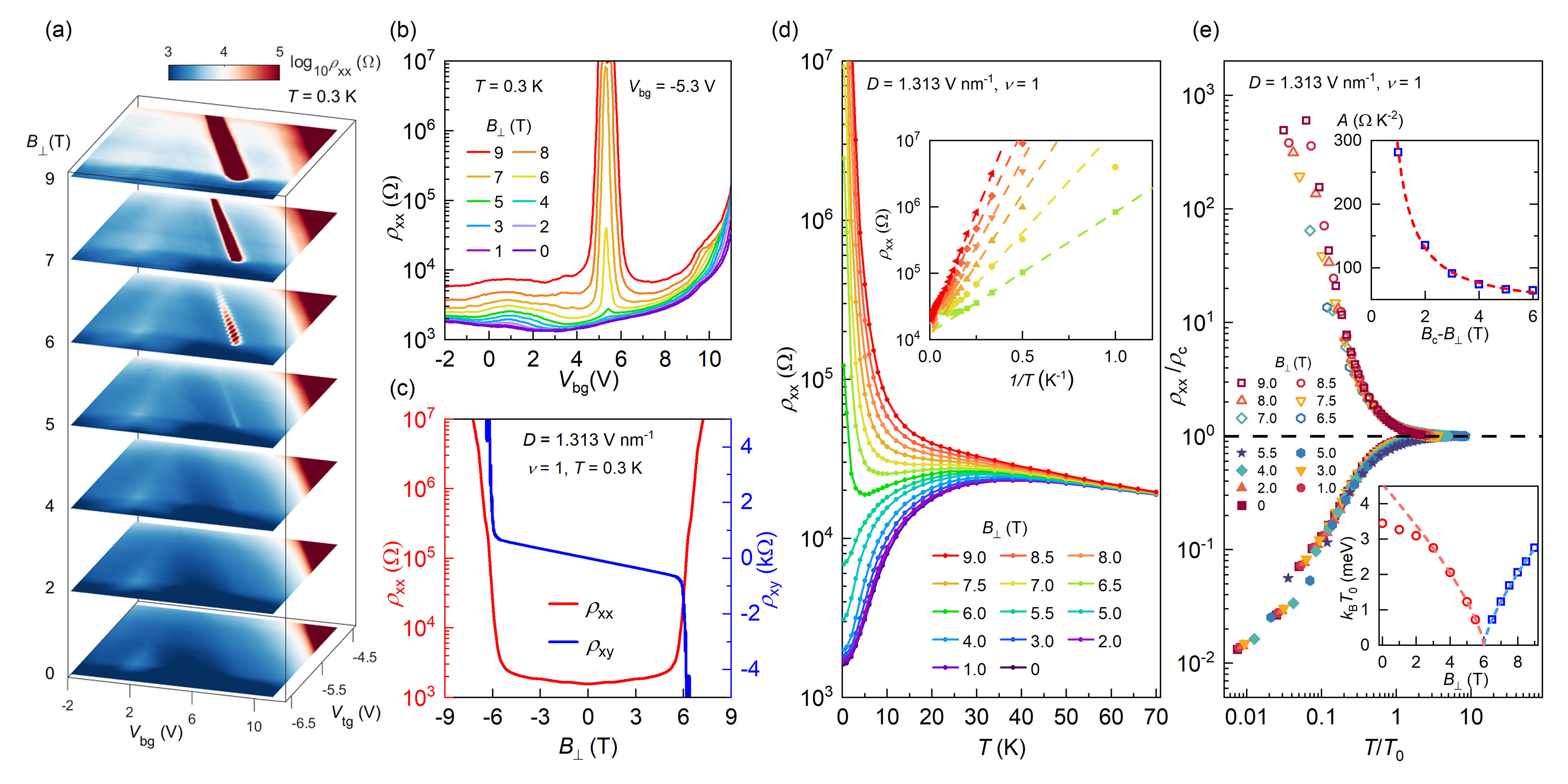}
	\caption{Magnetic-field-induced metal-insulator transition. (a) Evolution of the insulating state at half filling as a function of out-of-plane magnetic field $B_\perp$. Each color plane shows the $\rho_{xx}$ versus top and bottom gate voltages at a fixed $B_\perp$. (b) Showing $V_{bg}$ sweeps of $\rho_{xx}$ at a fixed $V_{tg}=-5.3$ V under varying $B_\perp$. Sharp resistance peaks occur at $\nu$ = 1 above the critical magnetic field $B_c$ $\approx$ 6 T. (c) Red line and blue line show $\rho_{xx}$ and $\rho$$_{xy}$ as functions of $B_\perp$ at 0.3 K, respectively. The gate voltages are fixed to ensure $\nu$  = 1 and $D$ = 1.313 V nm$^{-1}$, same for (d) and (e). (d) Temperature dependences of $\rho_{xx}$ at different $B_\perp$, showing the magnetic-field-induced MIT. Inset: Dashed lines represent Arrhenius fits to extract activation gap magnitudes for $B_\perp>6$ T. (e) Quantum critical scaling analysis of the MIT. The normalized resistance (by $\rho_{c}$ = $\rho_{xx}(B=B_c$)) curves neatly collapse onto two branches that are almost “mirror symmetric”, with one showing an insulating behavior and the other showing a metallic behavior. The determination of the scaling temperature $T_0$ is discussed in the main text. Top inset: magnetic field dependence of $A$, where $A$ is the fitting parameter for the low-temperature Fermi liquid. The dashed curve is a power law fit $A^{1/2}$ $\propto (B_c-B_{\perp})^{-0.98}$. Bottom inset shows the scaling parameter $k_BT_0$ versus $B_\perp$ on both sides of the critical field (red and blue open symbols). Close to $B_c$ ($|B_\perp-B_c|\le$ $\sim$3 T), both sides show power law dependences with critical exponents $\sim$0.73 (red and blue dashed curves).}
\end{figure*}
In the following, we demonstrate the quantum criticality induced by the out-of-plane magnetic field $B_{\perp}$ at $\nu$ = 1 ($D<D_t$). We plot the dual-gate mappings of $\rho_{xx}$ at different $B_{\perp}$ (Fig. 3a). A resistance peak at $\nu$ = 1 can be identified at $\sim$5 T for $D<D_t$ and quickly becomes prominent at higher fields (see linecuts in Fig. 3b). At fixed $D$ fields (see an example in Fig. 3c), the $\rho_{xx}$ weakly depends on $B_{\perp}$ at low magnetic fields, concomitant with a linear Hall resistance $\rho_{xy}$ due to the Lorentz force. The low-field Hall slope gives a density of $\sim$5.4$\times10^{12}$ cm$^{-2}$, close to the expected moiré density. Both $\rho_{xx}$ and $\rho_{xy}$ diverge at $\sim$6 T, which is then identified as the critical magnetic field $B_c$. The $B_{\perp}$ induced MIT at $\nu$ = 1 is clearly illustrated in Fig. 3d, where we show the temperature dependences of sheet resistance $\rho_{xx}$ (0.3 K to 70 K) at a few representative magnetic fields. Above $B_c$, the sample exhibits characteristic of an insulator following a thermal activation behavior $\rho_{xx}\propto E^{(\Delta/2k_{B}T)}$ (inset of Fig. 3d), where $k_{B}$ is the Boltzmann constant and $\Delta$ is the activation gap. At low temperature and below the critical field ($B_c$), the resistance drops upon cooling, exhibiting the characteristics of Fermi liquids. The extracted prefactor $A$ is plotted as a function of $|B_{\perp}-B_c|$ in the upper inset of Fig.3e, showing a power-law dependence $A^{1/2}\propto(B_c-B_{\perp})^{-0.98}$. Our result suggests a divergence of the effective mass ($\propto A^{1/2}$) at the MIT \cite{dobrosavljevic2012conductor, senthil2008theory}. After normalizing $\rho_{xx}(T)$ by the $\rho_{c}(T)$ at 6 T (critical field $B_c$) and scaling the temperature by a field-dependent $T_0$, the resistance curves readily collapse onto two branches in the Fig.3e log-log plot. On the insulating side ($B_{\perp}>B_c$), $T_0$ is calculated from the fitting parameter $\Delta/k_{B}$. While on the metallic side ($B_{\perp}<B_c$), $T_0$ is chosen to scale the curves nearly symmetric to the insulating branches about $\rho_{xx}/\rho_{c}$ = 1. The scaling parameter \emph{k$_{B}$T$_0$} continuously vanishes as it approaches the critical field (lower inset of Fig.3e). The $B_{\perp}$ dependences of $k_{B}T_0$ follows a power-law behavior $k_{B}T_0\propto |B_{\perp}-B_c|^{0.73}$ near both sides of the critical field (highlighted by the dashed red and blue curves, respectively). The critical exponent $\sim$0.73 is close to the value found in the bandwidth-tuned MIT reported by Li et al. \cite{li2021continuous}, indicating the quantum phase transitions tuned by the two different parameters share the same universality class.

\begin{figure*}[t]
	\centering
	\includegraphics[scale=1]{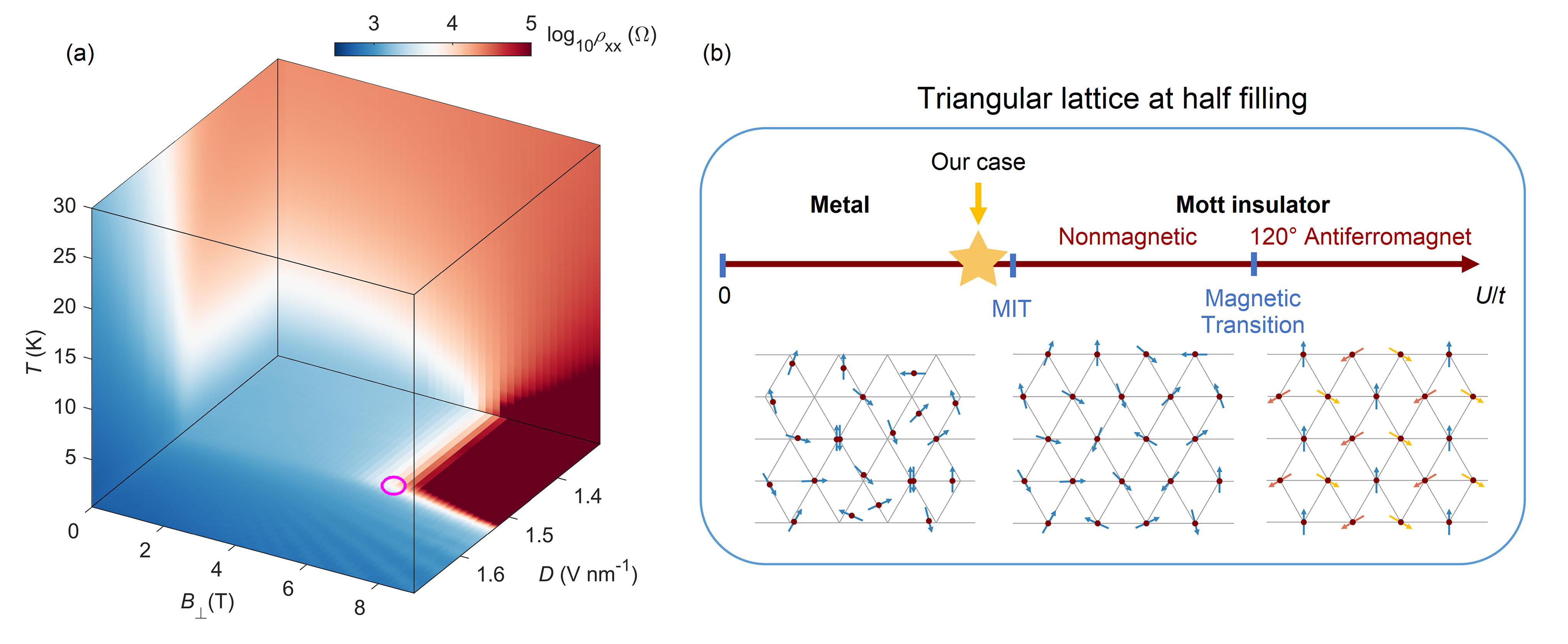}
	\caption{Phase diagram of 3L-MoTe$_2$/WSe$_2$ at half filling. (a) Longitudinal resistivity $\rho_{xx}$ (in log scale) as a function of $B$$_{\perp}$, $D$ and $T$. The diagram shows three slices of the function. Different states can be identified by the color contrast. The low-field light blue area denotes Fermi liquid phase, dark red area denotes correlated insulating phase. Lifshitz transition occurs near 1.5 V nm$^{-1}$, and generates a dark blue area at large $D$. In the dark blue area at the base temperature, Landau fan can be observed. The pink circle highlights the tricritical point in the $B$$_{\perp}$-$D$ plane. (b) The expected phase diagram of the triangular lattice Hubbard model. Our 3L-MoTe$_2$/WSe$_2$ lies on the left boundary of the MIT (denoted by the yellow arrow and star). Lower-panel schematics show the real-space charge distributions and spin orientations (denoted by the arrows) for corresponding phases. The metal and nonmagnetic insulator phases have random spin orientations while the system favors 120$^{\circ}$ Néel order in the large $U/t$ limit. }
\end{figure*}
Figure 4 summarizes our main observations at half filling of the 3L-MoTe$_2$/WSe$_2$ sample. In Fig. 4a, the resistance (in logarithmic scale) is plotted in the three-dimensional space projected on the $T$-$D$ ($B_{\perp}$ = 0), $T$-$B_{\perp}$ ($D$ = 1.313 V nm$^{-1}$) and $D$-$B_{\perp}$ ($T$ = 0.3 K) planes, respectively. First, the vertical electric field can induce charge transfer between different subbands and cause a breakdown of the single-band description of the system. It drives the system from a strongly interacting metal to a weakly interacting metal (evidenced by the great reduction of the $A$ value) accompanied by the Lifshitz transition. Secondly, the perpendicular magnetic field induces the metal-to-insulator quantum phase transition. A quantum tricritical point (highlighted by the pink circle) can be identified at $D_t$ = 1.49 V nm$^{-1}$ and $B_c$ = $\sim$6 T where phase transitions tuned by the two nearly independent parameters meet at the base temperature (see more discussions in Supplementary Figure 4).

In Fig. 4b, we illustrate the generally accepted phase diagram of the half-filled triangular lattice Hubbard model with tuning the interaction strength $U/t$. The ground state of our sample at $D<D_t$ and $B_{\perp}$ = 0 is a correlated metal exhibiting a Fermi-liquid behavior (with strongly renormalized effective mass) and Pomeranchuk effects. The observations are consistent with an intermediate coupling strength in the vicinity to the left side of the MIT \cite{wietek2021mott} (marked by the yellow star in Fig. 4b). The $U/t$ of our system should be smaller than the critical value of $(U/t)_c$ = 7 $\sim$ 8.5 according to different computational methods \cite{yoshioka2009quantum, yang2010effective, shirakawa2017ground, szasz2020chiral, wietek2021mott}. Spin fluctuations and geometric frustrations give rise to larger entropy in the exotic nonmagnetic insulating state (e.g. chiral QSL \cite{szasz2020chiral, wietek2021mott}) compared with the metal phase. Pomeranchuk effect arises due to the enhanced charge localization at higher temperatures since it yields a gain in free energy. It can also be seen from the Maxwell relation,
$$
	\left.\frac{\partial S}{\partial U}\right|_{T}=-\left.\frac{\partial F}{\partial T}\right|_{U}\text{,}
$$
where double occupancy (\emph{F}, denoting the fraction of the charges in doubly occupied lattice sites) decreases (namely single occupancy increases) as a function of temperature since the entropy \emph{S} is positively correlated with $U$ near the MIT \cite{wietek2021mott, georges1992numerical, werner2005interaction}. The perpendicular magnetic field driven MIT might originate from the orbital effects and reduction of bandwidth (and hence enhanced $U/t$) under $B_{\perp}$. Or it can be understood from another perspective as follows. Without any external fields, the sample is already charge gapped (evidenced by d$\rho_{xx}$/d$T<0$ and $\rho_{xx}>R_c$, see Fig. 1e) at $T>\sim$40 K. As the system is close to the MIT and the Fermi liquid has lower entropy, it becomes unstable towards metallicity upon cooling. When the excess spin fluctuations are suppressed by the magnetic field ($B_{\perp}>6$ T), entering the Fermi liquid state no longer saves entropy. The insulating (singly occupied) state naturally survives to the lowest temperatures (under $B_{\perp}$ field).

The vicinity to the intermediate coupling region in 3L-MoTe$_2$/WSe$_2$ provides a novel example of the richness and complex phase diagrams that can be unveiled in moiré superlattices. The observed Pomeranchuk effect is closely related to the higher entropy in the singly occupied state at half filling. Our observations could have implications on the existence of the elusive QSL phase at larger correlation strength, whereas compelling evidence is still missing. The multiband nature of the system also suggests it as a promising platform to construct and simulate the physics of a two-band Hubbard model \cite{zhang20214} or a moiré Kondo lattice \cite{kumar2021gate} with unprecedented controllability .

\begin{acknowledgments}
\end{acknowledgments}

\bibliography{references.bib}

\begin{appendix}
	\centering
	\includegraphics[page=1,scale=0.85]{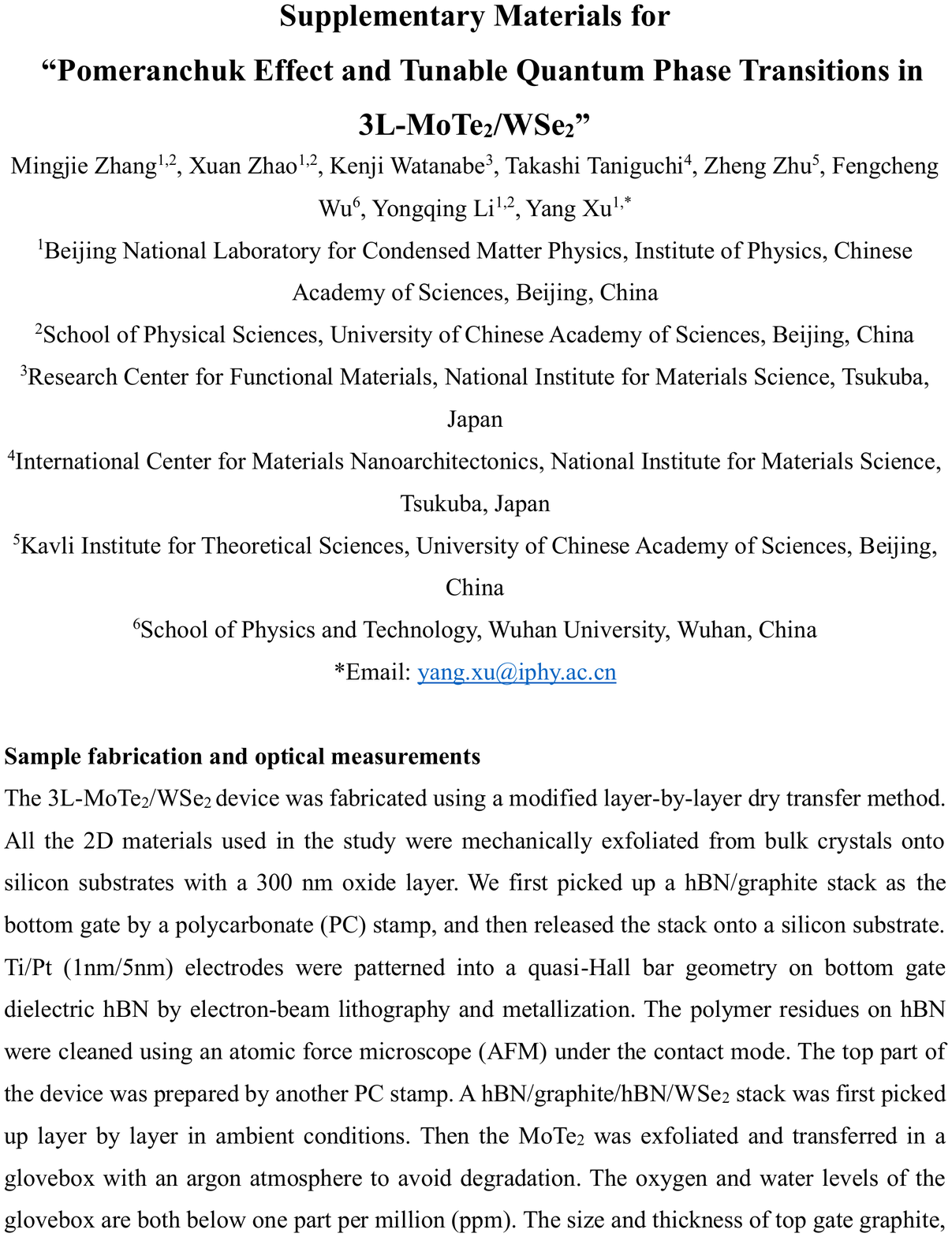} 
	\clearpage  
	\centering
	\includegraphics[page=2,scale=0.85]{Supplementary_information.pdf} 
	\clearpage
	\centering
	\includegraphics[page=3,scale=0.85]{Supplementary_information.pdf} 
	\clearpage
	\centering
	\includegraphics[page=4,scale=0.85]{Supplementary_information.pdf} 
	\clearpage
	\centering
	\includegraphics[page=5,scale=0.85]{Supplementary_information.pdf} 
	\clearpage
	\centering
	\includegraphics[page=6,scale=0.85]{Supplementary_information.pdf} 
	\clearpage
	\centering
	\includegraphics[page=7,scale=0.85]{Supplementary_information.pdf} 
	\clearpage
	\centering
	\includegraphics[page=8,scale=0.85]{Supplementary_information.pdf} 
	\clearpage

\end{appendix}

\end{document}